\documentclass[amstex,aps,preprint,eqsecnum]{revtex4}

\begin{document}

\title{A Note on Frame Dragging}

\author{R. F. O'Connell{\footnote{E-mail: oconnell@phys.lsu.edu \\
Phone: (225) 578-6848 Fax: (225) 578-5855
\\ }}}

\affiliation{Department of Physics and Astronomy, Louisiana State
University, \\ Baton Rouge, Louisiana  70803-4001, USA}

\date{\today}

\begin{abstract} 
The measurement of spin effects in general relativity has recently taken
centre stage with the successfully launched Gravity Probe B experiment
coming toward an end, coupled with recently reported measurements using
laser ranging.  Many accounts of these experiments have been in terms of
frame-dragging.  We point out that this terminology has given rise to
much confusion and that a better description is in terms of spin-orbit
and spin-spin effects.  In particular, we point out that the de Sitter
precession (which has been mesured to a high accuracy) is also a
frame-dragging effect and provides an accurate benchmark measurement of
spin-orbit effects which GPB needs to emulate.
\end{abstract}

\maketitle

\newpage

A measurement of the Lense - Thirring frame-dragging general relativistic
effect of the earth's rotation, by use of laser ranging to two earth
satellites [1] has recently been carried out, to within 10
original article, as well as  comments on same [2,3] assume that this is
the most accurate measurement so far of Òframe draggingÓ.  These articles
also discuss the de Sitter (geodesic) precession but refers to it as
being "--- one aspect of the class of relativistic phenomena loosely
known as gravitomagnetism--" [2], different from that of frame-dragging.
These are common beliefs, but they are not correct, as we now point out.
	
First, we point out that de Sitter precession is also a frame-dragging
effect.  In particular, because of its distance from the sun, the
earth-moon system can be regarded as a single body, which is rotating in
the gravitational field of the sun.  In other words, the earth-moon
system is essentially a gyroscope in the field of the sun (no different
in principle than the earth acting as a gyroscope or a quartz ball in the
Gravity Probe B (GPB) experiment acting as a gyroscope) and its
frame-dragging effect due to interaction with the sun has been measured,
using lunar laser ranging, to an accuracy of 0.35
accuracy which the GPB experiment or precession observations of the
recently discovered double pulsar system [6] needs to exceed!  We note
that the high accuracy achieved for the de Sitter geodetic effect
required analysis of gravitational three-body (earth, moon and sun)
theory but, to quote the authors of this work, "--geodetic precession is
implicit in the relativistic equations of motion -Ò [5]. Ultimately, we
expect that the most accurate results will emerge from analysis of the
double pulsar system (because the gravitational forces are significantly
stronger and there are no observational time limits, in contrast to the
GPB experiment), which has the added bonus of testing two-body effects.
	
Second, whereas "frame-dragging" is a very catchy appellation,
gravitational effects due to rotation (spin) are best described, using
the language of QED, as spin-orbit and spin-spin effects since they also
denote the interactions by which such effects are measured; in fact these
are the only such spin contributions to the basic Hamiltonian describing
the gravitational two-body system with arbitrary masses, spins and
quadruple moments) [7].  They manifest themselves in just two ways, spin
and orbital precessions and, whereas these can be measured in a variety
of ways (for example, as discussed in [8] orbital precession can be
subdivided into periastron, nodal and inclination precessions [9]), such
different measurements are simply "variations on the theme".

\newpage
	
Third, all measurements reported up to now as well as observations on
binary pulsar systems, have been confined solely to spin-orbit effects. 
It appears that the only immediate prospect of measuring the spin-spin
effect is the GPB experiment.  However, the significance of such a
measurement will be tempered by the fact that, given the verification of
the spin-orbit prediction, the predicted spin-spin result is necessary to
ensure conservation of total angular momentum [7].

\subparagraph{}

\textit{Note added proof.}  More recent analysis of lunar laser ranging 'increases
the uncertainty of the geodetic precession' [10].  Also, spin-orbit
coupling has been measured in the binary-pulsar system PSRB1534+12 [11].

\end{document}